\documentclass[conference]{IEEEtran}
\usepackage{graphicx}
\usepackage{flushend}
\usepackage{float}
\usepackage{amsthm}
\usepackage{amsmath}
\usepackage{comment}
\usepackage{cuted}
\usepackage{amssymb}
\usepackage{multirow}
\usepackage{mathtools, cuted}
\usepackage{lscape}
\usepackage[numbers,sort&compress]{natbib}
\usepackage{array}
\usepackage{algorithm}
\usepackage{xcolor}
\usepackage{algpseudocode}
\usepackage{multicol}
\usepackage{lettrine}
\usepackage{subcaption}

\usepackage[belowskip=-10pt,aboveskip=0pt]{caption}
\usepackage[margin=0.750in]{geometry}
\IEEEoverridecommandlockouts

\begin{document}

\flushend
\title{Physically-consistent Multi-band Massive MIMO Systems: A Radio Resource Management Model}
\author{Nuwan Balasuriya, Amine Mezghani, and Ekram Hossain}
\renewcommand\footnotemark{}
\thanks{The authors are with the Department of Electrical and Computer Engineering at the University of Manitoba, Canada. This work was partially supported by the project {\em 5G-Enabled Trustworthy Common Operational Picture with Edge Server Data Engine (5G-TCOP)} funded by Department of National Defense, Canada.}

\maketitle 
 
\begin{abstract}
Massive multiple-input multiple-output (mMIMO) antenna systems and inter-band carrier aggregation (CA)-enabled  multi-band communication are two key technologies to achieve very high data rates in beyond fifth generation (B5G) wireless systems. We propose a joint optimization framework for such systems where the mMIMO antenna spacing selection, precoder optimization, optimum sub-carrier selection and optimum power allocation are carried out simultaneously. We harness the bandwidth gain existing in a tightly coupled base station mMIMO antenna system to avoid sophisticated, non-practical antenna systems for multi-band operation. In particular, we analyze a multi-band communication system using a circuit-theoretic model to consider physical characteristics of a tightly coupled antenna array, and formulate a joint optimization problem to maximize the sum-rate. As part of the optimization, we also propose a novel block iterative water-filling-based sub-carrier selection and power allocation optimization algorithm for the multi-band mMIMO system. A novel sub-carrier windowing-based sub-carrier selection scheme is also proposed which considers the physical constraints (hardware limitation) at the mobile user devices. We carryout the optimizations in two ways:  (i) to optimize the antenna spacing selection in an offline manner, and (ii)  to select antenna elements from a dense array dynamically. Via computer simulations, we illustrate superior bandwidth gains present in the tightly-coupled colinear and rectangular planar antenna arrays, compared to the loosely-coupled or 
tightly-coupled parallel arrays. We further show the optimum sum-rate performance of the proposed optimization-based framework under various power allocation schemes and various user capability scenarios.
\end{abstract}


\section{Introduction}\label{Intro}
\lettrine{C}{arrier,} aggregation (CA) has been a popular technique used in fifth generation (5G) communication systems which allows different communication sessions to select better quality sub-carriers in a fading environment \cite{Jod21,Jod21_2,Nid22,Fu13,Kib20,Chi21}. Multi-band CA further extends this capability to select sub-carriers over multiple bands which in turn increases the spectral efficiency. Compared to single-band CA, multi-band CA can harness the diverse characteristics of the sub-carriers in different bands to produce a much diverse sub-carrier selection. For example, the high-band sub-carriers can have larger bandwidths than the low-band sub-carriers and also the high-band is less crowded as of today, hence have less interference\footnote{In this work we consider the spectrum proposed for beyond 5G systems and refer to sub-6GHz range as the low-band and 13GHz-20GHz range as high-band.}. However, the propagation and scattering properties of the high-band sub-carriers are poor and also larger bandwidth captures higher noise power. Thus, a combination of low and high-band sub-carriers would be beneficial for CA.

Meanwhile, massive multiple-input multiple-output (mMIMO) antenna systems have also been a very popular technique to achieve high spectral efficiency in modern communication systems. An extensive array of prior work on mMIMO has investigated the techniques to efficiently optimize the antenna selection, optimize beamforming and user scheduling in mMIMO systems. In  \cite{Ken22}, the authors optimize power allocation to the antennas and beamforming in mMIMO systems. In \cite{Ayu20}, the authors  optimize digital and analog beamforming for a mMIMO system while \cite{Ali19} optimizes a mMIMO system to obtain a trade-off between the energy efficiency and spectral efficiency.

A combination of multi-band CA and mMIMO can be expected to provide a superior spectral efficiency over the systems using multi-band CA or mMIMO alone. Although \cite{Bog15} deals with the optimization of CA-enabled mMIMO systems, it is limited to beamforming optimization. While \cite{Abd23} deals with joint carrier aggregation and mMIMO system optimization, it does not consider multi-band transmission. Also, the precoder optimization  is limited to zero-forcing precoding. In \cite{Bal24}, we proposed a joint optimization framework for multi-band mMIMO systems in which the optimization is carried out for precoding, sub-carrier selection, user selection and power allocation simultaneously. This communication system and the optimization shows a superior spectral efficiency over all the aforementioned systems.

However, the multi-band operation is always challenging and requires wide-band antenna systems. Two straight forward options for wide-band operation is to use multiple antennas/arrays to cater for different frequency band of interest (as used in the base station in \cite{Bal24}) or to have wide-band antennas which can resonate over multiple frequency bands simultaneously \cite{Kin18}. The former is not a very attractive solution as the cost of implementation and the form factor will increase significantly. In certain instances the regulations do not allow to install multiple sets of antennas. Especially at a miniaturized mobile device, it is not practical to deploy multiple antennas to cater for carriers in different frequency bands. The latter is an attractive solution and is indeed the practically used solution at the single antenna user devices \cite{Bal24}. 

Recently, it was demonstrated in \cite{Akr23} that a large multiple-input single-output (MISO) antenna array system is capable of providing a super wide-band response when the transmit antennas are sufficiently tightly coupled, even when the individual antennas are not specifically wide-band antennas. This bandwidth gain is very promising in multi-user multi-band communication systems where an adequately tightly coupled single transmit antenna array suffices to cater to multi-band communication.

To the best of our knowledge there is no joint optimization work on multi-band carrier aggregated mMIMO systems deploying tightly-coupled antenna arrays to harness the aforementioned bandwidth gain and use simple antennas in the antenna array. It is interesting to investigate the optimum tight coupling setup in the transmit antenna array to maximize the overall sum-rate under antenna array size constraints.
Furthermore, a joint optimization framework for multi-band mMIMO systems where the optimization is carried out for precoding, sub-carrier selection, user selection, power allocation, and transmit antenna spacing simultaneously would be required. 
 
 In this paper,  we propose a joint radio resource and transmit antenna system optimization framework for multi-band mMIMO systems. Our key contributions can be summarized as follows:
\begin{itemize}
\item Development of  a circuit-theoretic multi-user, multi-band communication system model to unify the wireless communication channel model and the antenna model (including mutual coupling effects).
\item Development of a joint optimization framework and the alternating simplification which produces a superior sum data rate over state-of-the-art frameworks. 
\item Solution of the optimization problem in two different ways, namely, the online (antenna spacing selection is carried-out dynamically together with other optimization parameters) and offline (antenna spacing is optimized only once, taking the average values for all the other optimization parameters, and then the other parameters are optimized dynamically for the set spacing value).
\item Investigation on the performance of the proposed framework under different power allocation schemes and user device constraints.
\end{itemize}

 The rest of the paper is organized as follows. In Section \ref{Sysmodel}, we describe the CA-enabled, circuit-theoretic multi-band mMIMO system model we use, followed by the descriptions of information flow from input to the output of the above model in Section \ref{Infoflow}. Using the information flow relationship, we develop the proposed joint optimization framework in Section \ref{Proposed}. In Section \ref{Complexity}, we conduct a complexity analysis of the proposed optimization approaches. Section \ref{Results} presents the simulation results to benchmark the sum-rate performance of the proposed joint optimization framework. Finally, Section \ref{Conc} concludes the paper.

\section{Basic System Model and Assumptions}\label{Sysmodel}

\subsection{CA-Enabled Multi-band mMIMO Transmission Model}\label{transmod}

\begin{figure}[!t]
\begin{center}
\includegraphics[width= 0.43\textwidth]{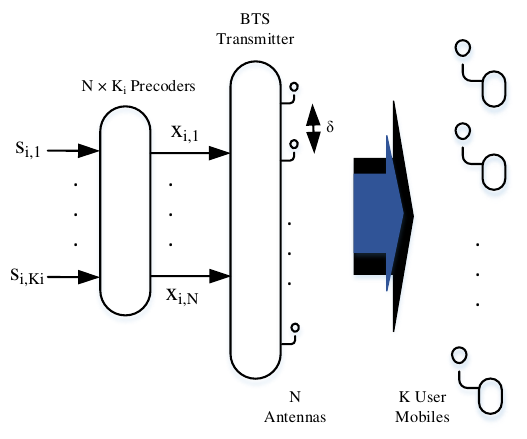}
\end{center}
\caption{Multi-band mMIMO communication system model.}
\label{model}
\end{figure}

\begin{figure}[!t]
\begin{center}
\includegraphics[width= 0.5\textwidth]{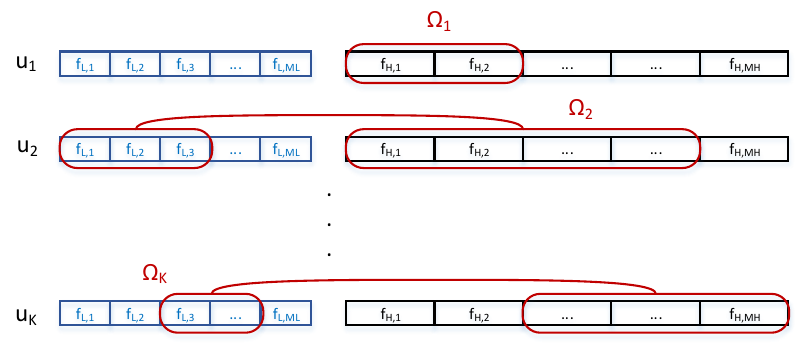}
\end{center}
\caption{Multi-band mMIMO resource allocation.}
\label{resapp}
\end{figure}

We consider a multi-band downlink mMIMO communication system shown in Fig.~\ref{model} comprising of a single base station (BS) having an $N$ antennas transmitting in low-band and high-band simultaneously. We consider a uniform $\delta$ spacing between the base station antennas which is a parameter to be optimized. However,  the maximum antenna array linear dimension is fixed at $D$ such that the number of antenna elements along one dimension in the transmit array is given by $N=\lfloor \frac{D}{\delta}\rfloor+1$. Note that the optimum number of antennas $N^*$ is decided by the optimum inter-element spacing $\delta^*$. Each of the $K$ ($\leq N$) user mobiles is equipped with a single antenna. We consider a set of independent $M_L$ and $M_H$ sub-carriers in the low-band and high-band, respectively, where the low-band sub-carriers have a bandwidth of $B_L$ while high-band sub-carriers have a bandwidth of $B_H$ in each.
Meanwhile,  each mobile user $k$, $k\in\{1,K\}$ can simultaneously operate in $\eta_k, (\eta_k\in\{1,2\})$ number of bands selected by switching the PIN diodes attached to the filters in the antenna system in the user devices which enables the power-constrained user devices  to operate only in one band at a time. 
Furthermore, each user $k$ is capable of handling a limited $\mathcal{N}_k$ number of sub-carriers selected contiguously within each band. The contiguous sub-carrier selection ensures high spectral efficiency and ease of implementation. We refer to the set of these sub-carriers, denoted by $\Omega_k$, as a {\em carrier window} (Fig.~\ref{resapp}). Note that $\{\eta_k,\mathcal{N}_k\}^{K}$ is known {\em a priori}.

\par The BS has a total power budget of $P_T$ for the downlink communication. Due to the difference in bandwidth of low-band and high-band sub-carriers, the additive white Gaussian noise (AWGN) variance in each band differs and these variances are given by $\sigma_L^2$ and  $\sigma_H^2(>\sigma_L^2)$, respectively. For simplicity of notation, let us denote the sub-carriers with a common index $i$ spanning over both low-band and high-band such that $f_i\in F=\{f_{L,1},f_{L,2},\dots,f_{L,M_L},f_{H,1},f_{H,2},\dots,f_{H,M_H}\}$. 

\subsection{Signal and Channel Model}\label{sigmod}

 In each sub-carrier $i$, the input signal vector $\textbf{s}_i=[s_{i,1}, s_{i,2}, \dots, s_{i,K_i}]^T$ is precoded using a $N \times K_i$ precoder matrix, $\textbf{W}^D_i$, which results in a signal vector $\textbf{x}_i=\textbf{W}^D_i\textbf{s}_i=[x_{i,1}, x_{i,2}, \dots, x_{i,N}]^T$. The signal is fed into the $N$ transmit antennas. 

The single antenna at a mobile device captures the superimposed signal modulated in the sub-carriers spanned over the two bands and delivers to the processing circuitry. We assume that each frequency sub-carrier undergoes independent quasi-static block Rayleigh fading where the fading coefficients are given by $\{\textbf{F}_i\}^{M_L+M_H}$, and the full knowledge of $\{\textbf{F}_i\}^{M_L+M_H}$ is available both at the receivers and the transmitter. At the  start of each block, $\{\textbf{F}_i\}^{M_L+M_H}$ is estimated and shared.

\subsection{Equivalent Circuit-Theoretic Model}\label{cctmodel}
\begin{figure*}[!t]
\begin{center}
\includegraphics[width= 0.85\textwidth]{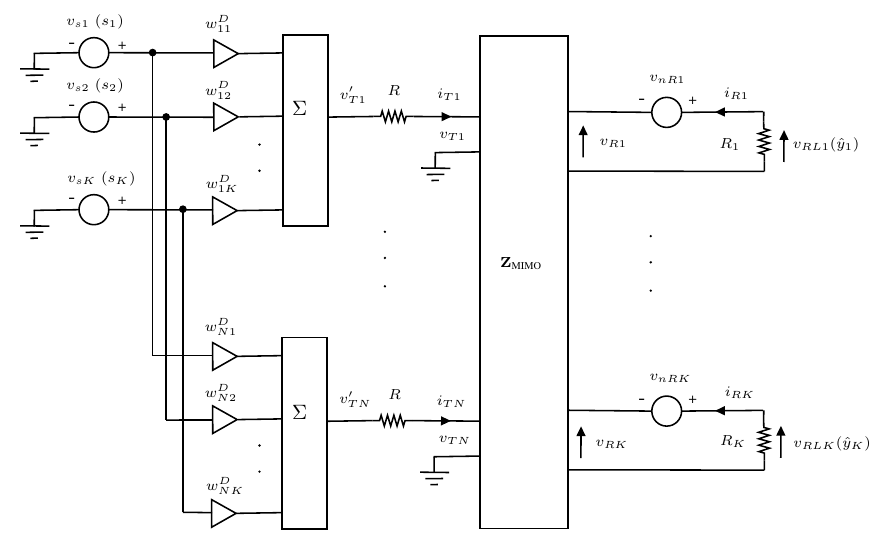}
\end{center}
\caption{Equivalent circuit-theoretic model for a mMIMO communication system  with a single sub-carrier.}
\label{cctmod}
\end{figure*}

Recall that in the considered communication system model, we propose to use a single transmit antenna array and single antenna at each of the users. Furthermore, we wish to harness the wideband characteristics resulting from tight coupling of transmit antennas rather than using sophisticated wideband antenna setups. Hence, the inter-element spacing needs to be optimized while considering the mutual antenna element coupling effects during the optimization. However, the mMIMO communication system defined in Sub-section \ref{transmod} defines a communication theoretic mMIMO channel while the antenna parameters lie in the domain of antenna theory. Inspired by \cite{Akr23}, we define an equivalent circuit theoretic unified model for both antennas and the mMIMO channel which is described as follows (Fig. \ref{cctmod}). Note that this model uses the multi-port network models defined in \cite{Ivr14} to jointly model the mMIMO channel and the antenna system.

We consider a set of binary generators $\textbf{v}_s=[v_{s1},v_{s2},\dots,v_{sK}]^T$ which represents the input signal vector and a set of amplifiers with gains given by the matrix $\mathbf{W}^D$. This amplifier operation is analogous to the conventional beamforming operation. This circuit operation results in a set of voltage signals $\textbf{v}'_{T}=[v'_{T1}, v'_{T2},\dots,v'_{TN}]^T$ which are the input voltage signals to the mMIMO antenna array. Let the input voltages, input currents, output voltages and output currents of the mMIMO system's equivalent multi-port circuit be denoted by $\textbf{v}_{T}=[v_{T1}, v_{T2},\dots,v_{TN}]^T$, $\textbf{i}_{T}=[i_{T1}, i_{T2},\dots,i_{TN}]^T$, $\textbf{v}_{R}=[v_{R1}, v_{R2},\dots,v_{RN}]^T$ and  $\textbf{i}_{R}=[i_{R1}, i_{R2},\dots,i_{RN}]^T$, respectively. $\textbf{v}_{nR}=[v_{nR1}, v_{nR2},\dots,v_{nRN}]^T$ denotes the AWGN added at the receiver antennas while $\textbf{v}_{RL}=[v_{RL1}, v_{RL2},\dots,v_{RLN}]^T$ denotes the load voltages as experienced at the mobile users having matched load impedances given by the $K\times K$ diagonal matrix $\mathbf{R}_L$ with real diagonal entries $\left[R_1, R_2, \dots, R_K\right]^T$. 
\par
The $\textbf{Z}_{\text{MIMO}}$ matrix completely defines the mMIMO channel and the antenna system, where
\begin{equation}\label{eqnMIMO}
\begin{bmatrix}
\textbf{v}_T(f) \\
\textbf{v}_R(f)
\end{bmatrix}
=
\underbrace{
\begin{bmatrix}
\textbf{Z}_T(f) & \textbf{Z}_{TR}(f) \\
\textbf{Z}_{RT}(f) & \textbf{Z}_R(f)
\end{bmatrix}
}_{\textbf{Z}_{\text{MIMO}}}
\begin{bmatrix}
\textbf{i}_T(f) \\
\textbf{i}_R(f)
\end{bmatrix}.
\end{equation}

Furthermore, the diagonal elements of $N \times N$-sized $\textbf{Z}_T(f)$ and $K \times K$-sized $\textbf{Z}_R(f)$ correspond to self-impedances of the transmit and receive array antennas, respectively, while the off-diagonal elements of $\textbf{Z}_T(f)$ represent the pair-wise mutual-impedances between tightly-coupled antenna elements in the transmit array. We consider receiver mobile devices each equipped with a single antenna and the separation between the mobile devices is very large. Hence, the mutual-impedances between any two mobile device antennas is negligible and we denote this by zero-valued off diagonal elements in $\textbf{Z}_R(f)$. Both $\textbf{Z}_T(f)$ and $\textbf{Z}_R(f)$ are symmetric matrices. $N \times K$-sized $\textbf{Z}_{TR}(f)$ and $K \times N$-sized $\textbf{Z}_{RT}(f)$ are transimpedance matrices representing the mMIMO channel, where $\textbf{Z}_{TR}(f)=\textbf{Z}_{RT}^T(f)$. However, in the far-field, the signal attenuation is very large, and due to no power input at the receiver, the effect of the receiver on the transmitter side is negligible. Thus, it is reasonable to represent this situation by $\textbf{Z}_{TR}(f) \approx \textbf{0}$.

\begin{figure}[!t]
\begin{center}
\includegraphics[width= 0.3\textwidth]{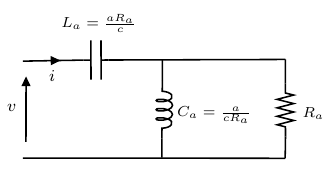}
\end{center}
\caption{Equivalent circuit for TM$_1$ antenna.}
\label{Chueqlnt}
\end{figure}

In this work, we assume all the antenna elements to be simple canonical minimum scattering (CMS) antennas given in \cite{Chu48}. Note that the lowest $Q$ factor (hence the maximum bandwidth) is obtained when the antenna operates in the lowest scattering mode, hence we consider the CMS antennas operating only in TM$_1$ lowest mode which is expected to provide the widest possible bandwidth response. We hereafter refer to the CMS antennas only radiating in TM$_1$ as {\em TM$_1$ antennas}. Most of the simple antennas such as dipole antennas can be approximated to TM$_1$ antennas. An equivalent circuit model for a TM$_1$ antenna is given in Fig. \ref{Chueqlnt} \cite{Chu48}, where the current and the voltage at the input port are given by
\begin{eqnarray}
i(f)&=&\sqrt{\frac{8\pi}{3}}\sqrt[4]{\frac{\mu}{\epsilon}}\frac{A_1}{k_0}\left[1+\frac{1}{jk_0a}\right]e^{-jk_0a}, \\ \nonumber
v(f)&=&\sqrt{\frac{8\pi}{3}}\sqrt[4]{\frac{\mu}{\epsilon}}\frac{A_1}{k_0}R_a\left[1+\frac{1}{jk_0a}-\frac{1}{(k_0a)^2}\right]e^{-jk_0a}, \nonumber
\end{eqnarray}
with amplitude $A_1$ and where, $k_0=\frac{2\pi f}{c}$ and $\mu, \epsilon, k_0, a$ and $c$  are the permeability, permittivity, TM$_1$ mode's complex coefficient, antenna element size (as in \cite{Chu48} an antenna which can be contained within a sphere having radius $a$), and the speed of light. Moreover, $L_a, C_a$, and $R_a$ represent the equivalent inductance, capacitance and the equivalent resistance in the TM$_1$ antenna model.

\begin{figure}[!t]
\begin{center}
\includegraphics[width= 0.2\textwidth]{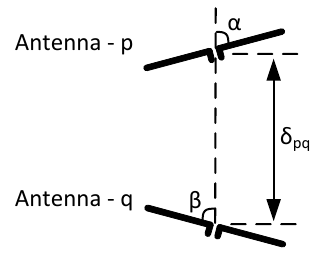}
\end{center}
\caption{Two TM$_1$ antennas arbitrary oriented in a 2-dimensional plane.}
\label{mutcup}
\end{figure}

\begin{figure}[!t]
\begin{center}
\includegraphics[width= 0.3\textwidth]{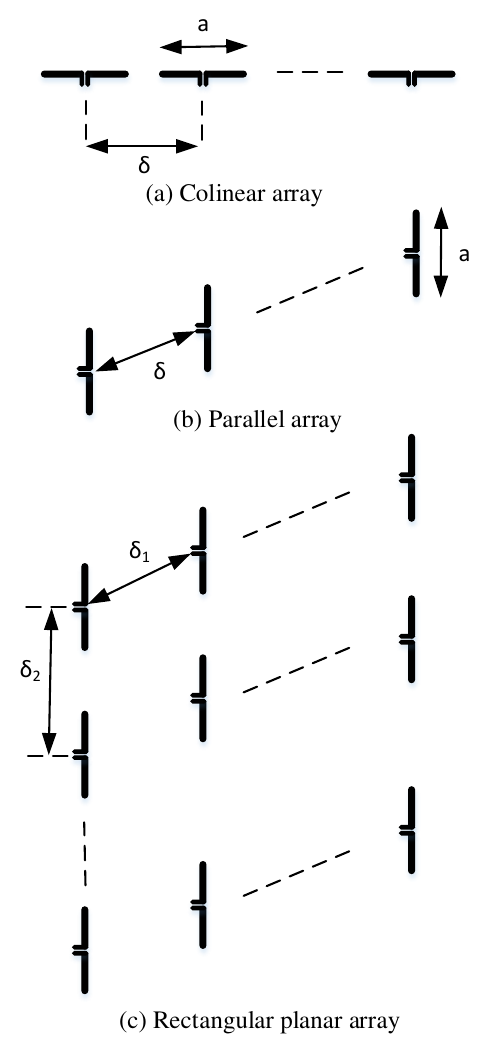}
\end{center}
\caption{Different array configurations.}
\label{arraytypes}
\end{figure}

Hence, the self-impedance of the antenna is given by
\begin{equation}\label{selfeq}
    Z(f)=\frac{v(f)}{i(f)}=\left[\frac{c^2+j2\pi fca-(2\pi fa)^2}{j2\pi fca-(2\pi fa)^2}\right]R_a,
\end{equation}
which can be used to determine the diagonal elements of $\textbf{Z}_T(f)$ and $\textbf{Z}_R(f)$.
Furthermore, the mutual-impedance between two TM$_1$ antennas $p$ and $q$ separated by a  distance $\delta_{pq}$ and with orientation angles $\alpha$ and $\beta$, as shown in Fig. \ref{mutcup}, is given by
\begin{equation}
\begin{split}
    Z_{p,q}(f)&=-3 \sqrt{\Re\{Z_p\} \Re\{Z_q\}}\times\\
    &\left[\frac{1}{2}\sin \alpha \sin\beta \left(\frac{1}{jk_0\delta_{pq}}+\frac{1}{(jk_0\delta_{pq})^2}+\frac{1}{(jk_0\delta_{pq})^3}\right)\right.\\
    & \left. +\cos \alpha \cos \beta \left(\frac{1}{(jk_0\delta_{pq})^2}
    +\frac{1}{(jk_0\delta_{pq})^3}\right)\right] e^{-jk_0\delta_{pq}},
\end{split}\label{mutualeq}
\end{equation}
where $\Re\{Z_p\}$ and $\Re\{Z_q\}$ refer to the real parts of the self-impedances of the two antennas. (\ref{mutualeq}) can be used to determine the off-diagonal elements of $\textbf{Z}_T$ \cite{Akr23_2}. In our work, we consider three different antenna element arrangements: \begin{enumerate}\item colinear array where $\alpha=\pi, \beta=0$, \item parallel array where $\alpha=\frac{\pi}{2}, \beta=\frac{\pi}{2}$, and \item $N_1 \times N_2$ rectangular planar array, which is a combination of colinear arrays and planar arrays (Fig.~\ref{arraytypes}).\end{enumerate}

The colinear configuration results in
\begin{equation}
\begin{split}
    Z_{p,q}(f)&=3 \sqrt{\Re\{Z_p\} \Re\{Z_q\}}\times\\
    &\left(\frac{1}{(jk_0\delta_{pq})^2}
    +\frac{1}{(jk_0\delta_{pq})^3}\right)e^{-jk_0\delta_{pq}},
\end{split}
\end{equation}\label{mutualeq1}
while the parallel configuration simplifies $Z_{p,q}(f)$ to
\begin{equation}
    \begin{split}
    Z_{p,q}(f)&=\frac{3}{2} \sqrt{\Re\{Z_p\} \Re\{Z_q\}}\times\\
    &\left(\frac{1}{jk_0\delta_{pq}}+\frac{1}{(jk_0\delta_{pq})^2}+\frac{1}{(jk_0\delta_{pq})^3}\right)
    e^{-j(k_0\delta_{pq}-\pi)}.
\end{split}\label{mutualeq2}
\end{equation}
Here, $\delta_{pq}=|p-q|\delta$.\par
In the rectangular planar array configuration, we find $\alpha=\pi-\beta$, hence the mutual-impedance simplifies to
\begin{equation}
\begin{split}
    Z_{p,q}(f)&=-3 \sqrt{\Re\{Z_p\} \Re\{Z_q\}}\times\\
    &\left[\frac{1}{2}\sin^2\beta \left(\frac{1}{jk_0\delta_{pq}}+\frac{1}{(jk_0\delta_{pq})^2}+\frac{1}{(jk_0\delta_{pq})^3}\right)\right.\\
    &\left.-\cos^2 \beta \left(\frac{1}{(jk_0\delta_{pq})^2}
    +\frac{1}{(jk_0\delta_{pq})^3}\right)\right] e^{-jk_0\delta_{pq}},
\end{split}\label{mutualeq3}
\end{equation}
where 
    $\delta_{pq}=\left\{\left[(p-q)+N_1\left(\lfloor\frac{q}{N_1}\rfloor -\lfloor\frac{p}{N_1}\rfloor\right) \right]^2\delta_1^2
    +\left[\lfloor\frac{p}{N_1}\rfloor \right.\right.$\\ 
    $\left.\left.-\lfloor\frac{q}{N_1}\rfloor\right]^2\delta_2^2\right\}^{\frac{1}{2}}$, $\beta=\tan^{-1}\left(\left|\frac{\lfloor\frac{p}{N_1}\rfloor-\lfloor\frac{q}{N_1}\rfloor}{(p-q)+N_1\left(\lfloor\frac{q}{N_1}\rfloor -\lfloor\frac{p}{N_1}\rfloor\right)}\right|\frac{\delta_2}{\delta_1}\right)$.
\par

We assume a Rayleigh faded wireless channels and we model the downlink channel by the transimpedance matrix. However, the mutual coupling at the transmitter antenna array affects the overall transimpedance matrix. A realization of the channel matrix under the influence of mutual coupling at the transmitter and receiver is given in \cite{Saa22} and we adopt the same Rayleigh fading model:
\begin{equation}
\begin{split}
       \textbf{H}(f)&=\frac{c}{2\pi f d^{\frac{\gamma}{2}}}\mbox{diag}(\Re\{\textbf{Z}_R(f)\})^{-\frac{1}{2}}\Re\{\textbf{Z}_R(f)\}^{\frac{1}{2}}\textbf{F}\\
       &~~~\times \Re\{\textbf{Z}_T(f)\}^{\frac{1}{2}}\mbox{diag}(\Re\{\textbf{Z}_T(f)\})^{-\frac{1}{2}},
\end{split}\label{channeleq}
\end{equation}
where $\textbf{F}$ is a matrix of size $K \times N$ with elements drawn from a complex Gaussian distribution with mean zero and variance $\frac{1}{2}$ per dimension, representing Rayleigh fading. Moreover, $d$ and $\gamma$ represent the distance between the transmit and receive antenna arrays in a far-field communication and the path-loss exponent, respectively.

Following the relationship between channel realization and the transimpedance given in \cite{Ivr14}, the effective transimpedance matrix can be determined as,
\begin{equation}
    \textbf{Z}_{RT}(f)=\frac{c}{2\pi f d^{\frac{\gamma}{2}}}\Re\{\textbf{Z}_R(f)\}^{\frac{1}{2}}\textbf{F}\Re\{\textbf{Z}_T(f)\}^{\frac{1}{2}}.\label{channeleq2}
\end{equation}
 Since the multi-user downlink consists of receiver antennas having negligible mutual coupling, (\ref{channeleq2}) simplifies to  
 \begin{equation}
    \textbf{Z}_{RT}(f)=\frac{c}{2\pi f d^{\frac{\gamma}{2}}}\Upsilon\textbf{F}\Re\{\textbf{Z}_T(f)\}^{\frac{1}{2}},\label{channeleq2_2}
\end{equation}
with $\Upsilon=\frac{2\pi fa}{\sqrt{(2\pi fa)^2+c^2}}\sqrt{R_a}$. Since we consider carriers in the GHz range, $\Upsilon\rightarrow \sqrt{R_a}$ when the receiver antennas are made sufficiently large to obtain matching.
\par With the knowledge of the full $\textbf{Z}_{\text{MIMO}}$ we can now derive a relationship between the input and output signals of the communication system taking the effect of mutual coupling  
into consideration.

\section{Input to Output Information Flow} \label{Infoflow}
Considering (\ref{eqnMIMO}) and the equivalent full circuit theoretic model shown in Fig. \ref{cctmod},
\begin{subequations}
\begin{equation}
\textbf{v}_T=\textbf{Z}_T\textbf{i}_T, \label{eq1}
\end{equation}
\begin{equation}
\textbf{v}_R=\textbf{Z}_{RT}\textbf{i}_T+\textbf{Z}_{R}\textbf{i}_R,  \label{eqn2}
\end{equation}
\begin{equation}
\textbf{v}_R=-\textbf{v}_{nR}-\textbf{R}_{L}\textbf{i}_R,  \label{eqn3}
\end{equation}
\begin{equation}
\textbf{v}_T=\textbf{v}'_{T}-R~\textbf{i}_T,  \label{eqn4}
\end{equation}
\begin{equation}
\textbf{v}_{RL}=-\textbf{R}_{L}\textbf{i}_R,  \label{eqn42}
\end{equation}
\begin{equation}
\textbf{v}'_T=\textbf{W}^D\textbf{v}_{s}.\label{eq5}
\end{equation}
\end{subequations}

Using (\ref{eq1}) and (\ref{eqn4}),
\begin{eqnarray}\label{eqA} 
    \textbf{Z}_Ti_T&=&\textbf{v}'_T-R~\textbf{i}_T, \nonumber \\ 
    \textbf{i}_T&=&(\textbf{Z}_T+R~\textbf{I})^{-1}\textbf{v}'_T.
\end{eqnarray}
Using (\ref{eqn2}) and (\ref{eqn3}) we obtain,
\begin{eqnarray}\label{eqB}  
    \textbf{Z}_{RT}\textbf{i}_T+\textbf{Z}_R\textbf{i}_R&=&-\textbf{v}_{nR}-\textbf{R}_L\textbf{i}_R, \nonumber\\ 
    (\textbf{Z}_R+\textbf{R}_L)\textbf{i}_R&=&-\textbf{v}_{nR}-\textbf{Z}_{RT}\textbf{i}_T. 
\end{eqnarray}
Applying $\textbf{i}_T$ of (\ref{eqA}) in (\ref{eqB}),
\begin{eqnarray} \label{A3}
(\textbf{Z}_R+\textbf{R}_L)\textbf{i}_R&=&-\textbf{v}_{nR}-\textbf{Z}_{RT}(\textbf{Z}_T+R~\textbf{I})^{-1}\textbf{v}'_T,\nonumber \\ 
\textbf{i}_R&=&-(\textbf{Z}_R+\textbf{R}_L)^{-1}\textbf{Z}_{RT}(\textbf{Z}_T+R~\textbf{I})^{-1}\textbf{v}'_T \nonumber\\~~~~~~~~&&-(\textbf{Z}_R+\textbf{R}_L)^{-1}\textbf{v}_{nR}.
\end{eqnarray}
By applying (\ref{eqn42}) and (\ref{eq5}) in (\ref{A3}) we obtain the relationship between $\textbf{v}_{RL}$ and $\textbf{v}_s$ given by
\begin{equation}
\begin{split} \textbf{v}_{RL}&=\textbf{R}_L\left(\textbf{R}_L+\textbf{Z}_R\right)^{-1}\textbf{Z}_{RT}(\textbf{Z}_T+R~\textbf{I})^{-1}\textbf{W}^D\textbf{v}_{s}\\ &~~~~~~~~~~~~+\textbf{R}_L\left(\textbf{R}_L+\textbf{Z}_R\right)^{-1}\textbf{v}_{nR}.
\end{split}\label{relationshipeq}
\end{equation}

(\ref{relationshipeq}) can be further represented in the form of a conventional wireless communication channel model as follows:
\begin{equation}
\textbf{v}_{RL}=\textbf{H}_{eq}\textbf{W}^D\textbf{v}_s+\textbf{v}_{n}, \label{eq6}
\end{equation}
with $\textbf{H}_{eq}=\textbf{R}_L\left(\textbf{R}_L+\textbf{Z}_R\right)^{-1}\textbf{Z}_{RT}(\textbf{Z}_T+R~\textbf{I})^{-1}$ and $\textbf{v}_{n}=\textbf{A}\textbf{v}_{nR}=\textbf{R}_L\left(\textbf{R}_L+\textbf{Z}_R\right)^{-1}\textbf{v}_{nR}$. Note that $\mathbf{A}$ is a diagonal matrix with non-zero diagonal entries, thus $\mathbf{A}$ is non-singular. Therefore, we can further normalize (\ref{eq6}) by multiplying both sides by $\mathbf{A}^{-1}$ to formulate
\begin{equation}
\tilde{\textbf{v}}_{RL}=\tilde{\textbf{H}}_{eq}\textbf{W}^D\textbf{v}_{s}+\textbf{v}_{nR}, \label{eq7}
\end{equation}
with $\tilde{\textbf{H}}_{eq}=\textbf{A}^{-1}\textbf{H}_{eq}$ and $\tilde{\textbf{v}}_{RL}=\textbf{A}^{-1}\textbf{v}_{RL}$. Note that $\tilde{\textbf{H}}_{eq}$ represents a modified channel coefficients matrix taking the effect of the mutual coupling in the antenna arrays into consideration.\\
\\ 
\textit{Result 1: The maximum radiation efficiency of the transmitter antenna system is given by the largest generalized eigenvalue of the pair} \{$\textbf{Z}_T,(\textbf{Z}_T+R~\textbf{I})$\}.\\

\textit{Proof:} Since we consider lossless antenna elements, the radiated power from the transmit antenna array is 
\begin{equation}
P_{Rad}=\textbf{i}^H_T\Re\{\textbf{Z}_T\}\textbf{i}_T.
\end{equation}
We also calculate the power loss in the transmitter as
\begin{equation}
P_{Loss}=R~\textbf{i}^H_T\textbf{i}_T.
\end{equation}
Thus, the radiation efficiency can be calculated as
\begin{equation}
\eta_{\mbox{eff}}=\frac{P_{Rad}}{P_{Loss}+P_{Rad}}=\frac{\textbf{i}^H_T\Re\{\textbf{Z}_T\}\textbf{i}_T}{\textbf{i}^H_T\left[\Re\{\textbf{Z}_T\}+R~\textbf{I}\right]\textbf{i}_T}.
\end{equation}
Note that both $\Re\{\textbf{Z}_T\}$ and $\Re\{\textbf{Z}_T\}+R~\textbf{I}$ are real symmetric matrices, hence $\frac{\textbf{i}^H_T\Re\{\textbf{Z}_T\}\textbf{i}_T}{\textbf{i}^H_T\left[\Re\{\textbf{Z}_T\}+R~\textbf{I}\right]\textbf{i}_T}$ is the generalized Rayleigh quotient. The maximum value of the Rayleigh quotient is equal to the largest root $\vartheta=\vartheta_{max}$ of the equation $\left|\Re\{\textbf{Z}_T\}-\vartheta\left[\Re\{\textbf{Z}_T\}+R\textbf{I}\right]\right|=0$ which proves {\em Result 1}.\\

\textit{Result 2: The mutual information carried in the output and input voltage signals} $\textbf{v}_{RL}$ \textit{and} $\textbf{v}_s$ \textit{in relationship (\ref{eq6}) is equal to the mutual information carried in the normalized output and input voltage signals} $\tilde{\textbf{v}}_{RL}$ and $\textbf{v}_s$ \textit{in relationship (\ref{eq7}).}\\

\textit{Proof:} Consider the relationship given in (\ref{eq6}). Mutual information,
\begin{eqnarray}
I(\textbf{v}_s;\textbf{v}_{RL})&=&\hbar\left(\textbf{v}_{RL}\right)-\hbar\left(\textbf{v}_{RL}|\textbf{v}_s\right)\\ \nonumber
&=&\hbar\left(\textbf{H}_{eq}\textbf{W}^D\textbf{v}_{s}+\textbf{v}_{n}\right)-\hbar\left(\textbf{v}_{n}\right)\\ \nonumber
&=&\log_2 \left|{\textbf{H}_{eq}\textbf{W}^D\textbf{W}^D}^H \textbf{H}^H_{eq}+\sigma^2\textbf{A}\textbf{A}^H\right|\\ \nonumber
&& ~~~~~~~~~~~~~~~~~-\log_2 \left|\sigma^2\textbf{A}\textbf{A}^H\right| \\ \nonumber
&=&\log_2 \left|\frac{1}{\sigma^2}\left(\textbf{A}^{-1}\textbf{H}_{eq}\textbf{W}^D{\textbf{W}^D}^H \textbf{H}^H_{eq}{\textbf{A}}^{-H}\right)+\textbf{I}\right|.
\end{eqnarray}
Similarly, considering (\ref{eq7}), the mutual information,
\begin{eqnarray}
I(\textbf{v}_s;\tilde{\textbf{v}}_{RL})&=&\log_2 \left|\tilde{\textbf{H}}_{eq}\textbf{W}^D{\textbf{W}^D}^H \tilde{\textbf{H}}^H_{eq}+\sigma^2\textbf{I}\right|\\ \nonumber
&& ~~~~~~~~~~~~~~~~~-\log_2 \left|\sigma^2\textbf{I}\right| \\ \nonumber
&=&\log_2 \left|\frac{1}{\sigma^2}\left(\tilde{\textbf{H}}_{eq}\textbf{W}^D{\textbf{W}^D}^H \tilde{\textbf{H}}^H_{eq}\right)+\textbf{I}\right|.
\end{eqnarray}
Using the relationship $\tilde{\textbf{H}}_{eq}=\textbf{A}^{-1}\textbf{H}_{eq}$,
\begin{eqnarray}\nonumber
I(\textbf{v}_s;\tilde{\textbf{v}}_{RL})&=&\log_2 \left|\frac{1}{\sigma^2}\left(\textbf{A}^{-1}\textbf{H}_{eq}\textbf{W}^D{\textbf{W}^D}^H \textbf{H}^H_{eq}{\textbf{A}}^{-H}\right)+\textbf{I}\right| \\ \nonumber
&\equiv&I(\textbf{v}_s;\textbf{v}_{RL}). \nonumber
\end{eqnarray}
Hence, when calculating the input-output information flow capacity in the downlink communication system in Fig. \ref{cctmod}, it is sufficient to calculate $I(\textbf{v}_s;\tilde{\textbf{v}}_{RL})$. 

\section{Proposed Resource Management Framework} \label{Proposed}

Our aim is to determine the optimal BS antenna element spacing $\delta$, the set of optimum precoding matrices $\{\textbf{W}^D_i\}, i\in F$, the optimal power allocation for different users $\{P_{i,k}\}$, the optimum sub-carrier set (together with the set of window position $\{\Omega_k\}$), and the maximum sum-rate via joint optimization. For this purpose, we use the circuit-theoretic model we introduced in Subsection \ref{cctmodel}, and in order to maximize the sum-rate, we maximize the sum of mutual information between the input voltage signal $\textbf{v}_s$ and individual single antenna users received signal $v_{RLk}$ for each of the $K$ users. With {\em Result 2}, it is equivalent to maximizing the sum of mutual information between the input voltage signal $\textbf{v}_s$ and normalized individual single antenna users received signal $\tilde{v}_{RLk}$ for each of the $K$ users.
 
\subsection{Problem Formulation}

Let us consider a $K_i \times N$ equivalent channel coefficients matrix for each sub-carrier $i$ given by $\tilde{\textbf{H}}_{{eq}_i}$  where the rows of $\tilde{\textbf{H}}_{{eq}_i}$  are given by $\tilde{\textbf{h}}_{i,k}, ~k\in[1,K_i]$ representing the equivalent (mutual coupling effects incorporated) channel coefficient vector from BS antenna array to $k$-th user's antenna. The equivalent multi-user mMIMO system possesses the uplink downlink dual capacity relationship \cite{Hea19}, thus, the optimum downlink sum-rate can be determined considering an uplink with channel coefficient matrix $\tilde{\textbf{H}}^H_{{eq}_i}$, AWGN noise with zero mean and $\sigma^2$ variance per dimension, added at each BS antenna and the same sum power at the transmitters. Furthermore, let $\textbf{w}_{i,k}$ represent the $k$-th row of the dual uplink precoding matrix $\textbf{W}_i$. Under minimum mean square error-successive interference cancellation (MMSE-SIC) detection, multi-user mMIMO uplink sum-rate per unit bandwidth in sub-carrier $i$ can be calculated as in \cite{Hea19}, and is given by $\log_2 \left|\mathbf{I}+\sum_{k=1}^{K_i}\frac{\left(\tilde{\textbf{h}}^H_{i,k}\textbf{w}_{i,k}\textbf{w}^H_{i,k}\tilde{\textbf{h}}_{i,k}\right)}{\sigma_i^2}\right|$.
Hence, the joint optimization problem can be stated as follows:
\begin{subequations}\label{first:main}
\begin{equation}\tag{\ref{first:main}}
\max\limits_{\substack{\delta_1,\delta_2, P_{i,k},\\ \textbf{w}_{i,k}, \Omega_k}}~R=\sum\limits_{i=1}^{M_L+M_H} B_i \log_2 \left|\mathbf{I}+\\ \sum_{k=1}^{K_i}\frac{\left(\tilde{\textbf{h}}^H_{i,k}\textbf{w}_{i,k}\textbf{w}^H_{i,k}\tilde{\textbf{h}}_{i,k}\right)}{\sigma_i^2}\right|
\end{equation}
~~~~~~~~~~~~~~~~~s.t.
\begin{equation}\label{first:a}
P_{i,k}=tr (\textbf{w}_{i,k}\textbf{w}^H_{i,k})
\end{equation}
\begin{equation}\label{first:a2}
\tilde{\textbf{h}}_{i,k}=\boldsymbol{f}_{i,k}(\delta_1,\delta_2)
\end{equation}
\begin{equation}\label{first:b}
\sum_{i=1}^{M_L+M_H}\sum_{k=1}^{K_i}P_{i,k}\leq P_T,~~~ P_{i,k}\geq0
\end{equation}
\begin{equation}\label{first:d}
P_{i,k}=0~~~\mbox{for}~~~f_i\notin\Omega_k
\end{equation}
\begin{equation}\label{first:e}
(N_1-1)\delta_1\leq D_1~~\mbox{and}~~(N_2-1)\delta_2+a\leq D_2.
\end{equation}
\end{subequations}

Constraints  (\ref{first:a2}) and (\ref{first:e}) are valid for a rectangular planar array with horizontal and vertical dimensions given by $D_1$ and $D_2$. In the case of linear antenna arrays, constraint (\ref{first:a2}) has to be replaced by $\tilde{\textbf{h}}_{i,k}=\boldsymbol{f}_{i,k}(\delta)$ and (\ref{first:e}) by only one of the inequalities.
$P_{i,k}$ is the power allocated to $k$-th user's communication over sub-carrier $i$. The constraint (\ref{first:b}) on power allocation can be replaced with $\sum_{j=1}^{M_L}\sum_{k=1}^{K_j}P_{j,k}\leq \frac{1}{1+\beta}P_T$ and $\sum_{i=1}^{M_H}\sum_{k=1}^{K_i}P_{i,k}\leq \frac{\beta}{1+\beta}P_T$, $0\leq \beta < \infty$ when band-wise power allocation constraints are implemented while it can be replaced with $P_{i,k}\leq \frac{P_T}{M_L+M_H}$, when sub-carrier-wise power constraints are implemented.

\subsection{Inner and outer Optimization Sub-problems}
It is apparent that this optimization problem is a non-convex problem when a joint optimization is considered for $\{\delta_1, \delta_2\}$ (contained within ${\textbf{h}}_{i,k}$), and $P_{i,k}$, $\textbf{w}_{i,k}$ together with $\Omega_k$. Hence, we divide the full optimization problem into two sub problems:
\begin{enumerate}
    \item Optimization for $\mathbf{\delta}=\{\delta_1,\delta_2\}$
    \item Optimization for $P_{i,k}$, $\textbf{w}_{i,k}$ and $\Omega_k$.
\end{enumerate}
Note that a parallel alternating optimization is infeasible in this particular problem as the dimensions of the equivalent channel gain vectors and the precoding vectors are dependent on the value of $\delta$. Hence, we consider a nested optimization approach as described below.\par
For a given $\mathbf{\delta}$ value set preserving the validity of constraint (\ref{first:e}), corresponding ${\textbf{h}}_{i,k}$ vectors can be calculated using (\ref{selfeq})-(\ref{channeleq2}) and (\ref{relationshipeq}). Then, for the calculated ${\textbf{h}}_{i,k}$, we can carryout a joint optimization for $P_{i,k}$, $\textbf{w}_{i,k}$ and $\Omega_k$. The resultant optimal sum-rate value represents a single point in a one-dimensional function of optimum sum-rate versus independent variable $\mathbf{\delta}$ (two dimensional function in the case of planar arrays). Now our problem converts to the optimization of this single dimensional optimal sum-rate function with respect to $\delta$ (two dimensional optimal sum-rate function with respect to $\mathbf{\delta}=\{\delta_1,\delta_2\}$) and finding the maximum value of optimum sum-rate when $\mathbf{\delta}$ is varied.
Let us denote the optimization sub-problem where the $\delta$ is optimized as {\em outer optimization} and the optimization of other variables for a known $\mathbf{\delta}$ as {\em inner optimization}. 


\subsection{Simplified Alternating Inner Optimization}\label{inner}
The inner optimization problem (i.e., the constrained optimization problem in (\ref{first:main}) for a given $\delta$, hence for a given set of $\textbf{h}_{i,k}$) can now be converted to an alternating optimization as shown in \textbf{Algorithm~\ref{inneroptalg}} \cite{Bal24}. Here, $R_{\mbox{prev}}~\mbox{and} ~R_{\mbox{new}}$ refer to the sum-rate calculated at the end of the previous iteration and sum-rate calculated at the current iteration.

\begin{algorithm}
\caption{Alternating inner optimization algorithm \cite{Bal24}.}\label{inneroptalg}
\begin{algorithmic}[1]
\State Calculate $\{\textbf{H}_i\}$ for input $\delta$
\State Initialize precoders, $\{\textbf{W}_i\} \gets \{\textbf{W}_i^{\mbox{init}}\}$
\Repeat
\State Allocate sub-carriers to users
\State Allocate power to user-sub-carrier blocks
\State Calculate precoders, $\{\textbf{W}_i\}_{\mbox{new}}$
\State Update precoders,~$\{\textbf{W}_i\} \gets \{\textbf{W}_i\}_{\mbox{new}}$
\State Calculate sum-rate $R_{\mbox{new}}=\sum_{\forall k} R_{k,\mbox{new}}$
\Until{$|R_{\mbox{prev}}-R_{\mbox{new}}|< \Delta R_{\mbox{Th}}$}
\end{algorithmic}
\end{algorithm}

Note that the sub-carrier allocation and power allocation here in steps 5 and 6 are carried out using an iterative water filling approach discussed in Subsection \ref{precopt}. 

\subsection{Block Iterative Water-filling Joint Precoding and Power Allocation}\label{precopt}
One of the important components of the proposed joint inner optimization algorithm is the iterative water-filling-based power and sub-carrier allocation algorithm. 
We adopt the iterative water-filling algorithm presented in \cite{Wei04} and modify the same to fit the multi-band systems \cite{Bal24}. \par
As in \cite{Wei04}, the dual uplink sub-carrier and power allocation optimization problem can be simplified as:
\begin{small} 
\begin{eqnarray}
\arg~\max_{P_{i,k},\textbf{w}_{i,k}}\sum_{i=1}^{M_L+M_H} B_i \log_2\left|\textbf{Z}_{i,k}+\left(\tilde{\textbf{h}}^H_{i,k}\textbf{w}_{i,k}\textbf{w}^H_{i,k}\tilde{\textbf{h}}_{i,k}\right)\right|\\ \nonumber
-B_i\log_2\left|\textbf{Z}_{i,k}\right|,
\end{eqnarray}
\end{small}
where~~$\textbf{Z}_{i,k}=\sum_{k'=1,k'\neq k}^{K_i}\tilde{\textbf{h}}^H_{i,k'}\textbf{w}_{i,k'}\textbf{w}^H_{i,k'}\tilde{\textbf{h}}_{i,k'}+\sigma^2_i\mathbf{I}$ is the effective interference and noise variance as seen by the BS in dual-uplink corresponding to $k$-th user communication, considering all the other communications intended for other users as interference.
Since $\textbf{Z}_{i,k}$ is a positive semi-definite matrix, $\textbf{Z}_{i,k}$ can be expressed as $\textbf{Q}_{i,k}\mathbf{\Delta}_{i,k} \textbf{Q}_{i,k}^H$, where $\textbf{Q}_{i,k}$ is a unitary matrix and $\mathbf{\Delta}_{i,k}$ is a diagonal matrix with eigenvalues. Thus, we can define $\dot{\textbf{h}}_{i,k}=\tilde{\textbf{h}}_{i,k}\textbf{Q}_{i,k}\mathbf{\Delta}_{i,k}^{-\frac{1}{2}}$ such that the effect of noise and interference is now absorbed to the channel gain matrix. Now the optimization can be written as:
\begin{equation} \label{eq3}
\arg~\max_{P_{i,k}, \textbf{w}_{i,k}}\sum_{i=1}^{M_L+M_H} B_i \log_2\left|\mathbf{I}+\left(\dot{\textbf{h}}^H_{i,k}\textbf{w}_{i,k}\textbf{w}^H_{i,k}\dot{\textbf{h}}_{i,k}\right)\right|.
\end{equation}

The precoder optimization and power allocation is carried out iteratively as follows. First, we initialize the precoder vector set in carrier $i$, given by $\textbf{W}_i=[\textbf{w}_{i,1},\textbf{w}_{i,2},\dots,\textbf{w}_{i,K_i}]$, $i\in\{1,2,\dots,M_L+M_H\}$. Then we calculate the sum interference from all other users and the noise in sub-carrier $i$, corresponding to user $k$, $\textbf{Z}_{i,k}$ and the noise and interference absorbed channel coefficient vector, $\dot{\textbf{h}}_{i,k}$. Let the single non-zero eigenvalue correspond to $\dot{\textbf{h}}_{i,k}\dot{\textbf{h}}^H_{i,k}$  be $\Lambda_{i,k}$ ($\Lambda_{i,k}=\left\|\dot{\textbf{h}}_{i,k}\right\|^2$). Similarly, we can calculate the corresponding eigenvalues and the equivalent interference and noise at each of the users in each of the sub-carriers. Let $\textbf{U}_{i,k}\Lambda^{\frac{1}{2}}_{i,k} \textbf{V}_{i,k}^H$ and $\textbf{S}_{i,k}\Sigma^{\frac{1}{2}}_{i,k} \textbf{T}_{i,k}^H$ be the singular value decompositions of $\dot{\textbf{h}}_{i,k}$ and $\textbf{w}_{i,k}$, respectively, where  $\textbf{U}_{i,k},\textbf{V}_{i,k},\textbf{S}_{i,k},\textbf{T}_{i,k}$ are unitary matrices each. We can maximize the objective function in (\ref{eq3}) by selecting $\textbf{S}_{i,k}=\textbf{U}_{i,k}$, hence we can rewrite the optimization problem as:
\begin{equation}
\max_{P_{i,k}}R=\sum_{i=1}^{M_L+M_H} B_i \sum_{k=1}^{K_i}\log_2\left(1+\Lambda_{i,k}P_{i,k}\right).
\end{equation}
Now we are in a position to consider the system as a parallel set of channels communicating independently and allocate power using a standard water-filling algorithm where the optimum power allocation to user $k$ in $i$-th carrier is given by $P^{{opt}}_{i,k}=\left[\kappa_{i}-\frac{1}{\Lambda_{i,k}}\right]^+$ such that $\sum_{i=1}^{M_L+M_H}\sum_{k=1}^{K}P^{opt}_{i,k}\leq P_T$. $[\alpha]^+$ denotes a function whose output value is $\alpha$ if $\alpha \geq 0$ and $0$ when $\alpha<0$. Here, $\kappa_{i}$ can take two different values based on the band the carrier $i$ belongs to. However, the ratio between the $\kappa_{i}$ values for high-band to low-band sub-carriers, $\epsilon$ is the same as the bandwidth ratio of the sub-carriers. Hence, the total sum-rate can be determined as  $R^{opt}=\sum_{i=1}^{M_L+M_H}\sum_{k=1}^{K_i}B_i\left[\log_2\left(\Lambda_{i,k}\kappa_{i}\right)\right]^+$. Furthermore, we can calculate the corresponding precoding vector for user $k$ in $i$-th sub-carrier using $\textbf{w}_{i,k}=\textbf{U}_{i,k}[P^{\frac{1}{2}}_{i,k}~\textbf{0}]^T$.
Note that, at the end of the iteration, we calculate a set of updated uplink precoder matrices which we treat as the starting uplink precoder set for the next iteration. This iterative precoder optimization is carried out until the sum-rate does not show a significant further improvement. 
Also note that, this sub-optimal precoder optimization allocates zero power to some user-carrier combinations so that some sub-carriers can be fully unused by any of the users. Thus, the iterative water-filling sub-algorithm produces an effective joint sub-carrier selection and precoder optimization. It is worth mentioning that the dual uplink precoder $\textbf{W}_i$ is different from the downlink precoder $\textbf{W}^D_{i}$. However, $\textbf{W}^D_{i}$ can be determined via a linear transformation of $\textbf{W}_{i}$. For brevity, we omit the details in this paper (see \cite{Hea19}).

\subsection{Selecting Windowed Channels}\label{window}
We use the same subcarrier windowing technique as in \cite{Bal24}, which can be described as follows. As mentioned in Section \ref{Sysmodel}, some devices are capable of supporting only a limited set of sub-carriers at a time. During the joint precoder optimization and sub-carrier selection step described in Subsection \ref{precopt} we also accommodate a sub-carrier selection constraint to represent the aforementioned device limitations. While selecting sub-carriers for power allocation, we also optimize the carrier windows vector $\mathbf{\Omega}=\{\Omega_1,\dots,\Omega_K\}$. One exhaustive approach to perform this $\mathbf{\Omega}$ optimization is to run the joint inner optimization algorithm for each and every possible carrier window set combination satisfying $\{\eta_k,\mathcal{N}_k\}^{K}$. However, this approach requires us to determine the sum-rates for a total of $\prod_{k=1}^{K} \phi_k$ different combinations, where $\phi_k$ is the total number possibilities for $\Omega_k$. Therefore, we propose a simplified approach to determine the optimal $\mathbf{\Omega}$. Recall that during each iteration we calculate the value $\Lambda_{i,k}$ which is a good measure of how much power is allocated to each user in each sub-carrier. A large $\Lambda_{i,k}$ indicates a large power allocation and vise-versa. Furthermore, recall that $\kappa_i$ is different for low-band and high-band sub-carriers, and as a result, there is an $\epsilon$ times bias for power allocation for high-band sub-carriers than for low-band sub-carriers. Thus, for each user $k$ we select the $\Omega_k$ such that $\sum\bar{\epsilon}\Lambda_{i,k}$ is a maximum, where $\bar{\epsilon}$ is $\epsilon$ for high-band sub-carriers and $1$ for low-band sub-carriers. This approach reduces the complexity of the search for the optimum window to $\sum_{k=1}^{K}\phi_k$. Power allocation considers only sub-carriers in the selected windows only.

\subsection{Outer Optimization and the Overall Algorithm}\label{outer}
During the outer optimization we maximize the sum-rate function (let us denote this function by $g(\delta$)) with respect to $\delta$ under the constraint (\ref{first:e}). However, the function itself is non-convex, hence we follow a global optimization approach for the outer optimization. We primarily deploy {\em Particle Swarm Optimization} algorithm \cite{Cho13} to find the global maximum. In comparison to a conventional optimization, particle swarm optimization simultaneously initiates search from a multiple randomly located points over the domain of the objective function which are known as particles. Moreover, each particle is assigned a random velocity which represents the direction and the amount of movement in the search space. Then, during each iteration, the velocity is updated such that these particles are pushed towards a global optimum. Due to the fact that the multiple particles are optimized simultaneously, the algorithm converges to the global minimum with high probability. However, particle swarm algorithm requires a very large number of iterations for convergence, hence a large number of function evaluations within each of the iterations. Recall that each of these function evaluations requires to run the \textbf{Algorithm~\ref{inneroptalg}} which itself is a complex operation. Thus, it is prohibitively expensive to resort only to a particle swarm based optimization for the outer optimization. Here, to have an affordable complexity we use several iterations of the particle swarm optimization to bring the search to a point within the neighbourhood of the global maximum and then follow a Swan bracketing and Golden rule bisection-based line search \cite{Cho13} to find the global maximum. In the case of planar antenna arrays, with the presence of two optimization variables we replace the line search with the Gradient Ascent Algorithm \cite{Cho13} with a stopping criteria of reaching a minimum given sum-rate improvement per iteration.

 The overall optimization algorithm can be stated as shown in \textbf{Algorithm \ref{outeroptalg}}\footnote{$g(.)$ requires a single run of inner optimization given in \textbf{Algorithm \ref{inneroptalg}}, $\tilde{\delta}$ and $\dot{\delta}$ are two predefined threshold constants.}.

\begin{algorithm*}
\caption{Overall optimization algorithm.}\label{outeroptalg}
\begin{algorithmic}[1]
\State Constants: $C_1=0.8,C_2=2,C_3=2,\zeta=10$ (number of particles), $\mathcal{I}_{PS}$ (number of iterations for particle swarm)\Comment{Particle swarm optimization}
\State Select random particle points in the domain of $g(.)$: $\{\textbf{x}\}^{\zeta}$
\State Select random velocity vectors in $[-\frac{1}{2},\frac{1}{2}]: \textbf{v}^{\zeta}$
\State $\{\textbf{pbest}\}^{\zeta}\gets\{\mathbf{\delta}\}^{\zeta}$
\State $gbest \gets \max\{\textbf{pbest}\}$
\For{$t=1:\mathcal{I}_{PS}$}
\State Generate random vectors $\textbf{r}, \textbf{s}$ in the domain $[0,1]^\zeta$
\State Update: $\textbf{v} \gets C_1\textbf{v}+C_2\textbf{r}\odot(\textbf{pbest}-\textbf{x})+C_3\textbf{s}\odot(gbest.[\textbf{1}]^{\zeta}-\textbf{x})$
\State Update: $\textbf{x} \gets \textbf{x}+\textbf{v}$
\If{Any of the new pints in \textbf{x} are outside the range of g(.)}
\State Map the outlier points to the domain margins
\EndIf

\If{$g(x)$ for any new point in $x\in\textbf{x}$ is greater than $g(.)$ for corresponding element of $\textbf{pbest}$}
\State Update \textbf{pbest}
\EndIf
\State $gbest \gets \max\{\textbf{pbest}\}$
\State $\delta_0 \gets gbest$ \Comment{End of particle swarm optimization}
\EndFor
\vskip 0.2cm
\State Initialize, $\delta_L \gets \delta_0$, $\delta_R \gets 
\delta_0+\tilde{\delta}$ (Initial interval: $[\delta_L,\delta_R]$)\Comment{Swan bracketing}
\State Swan bracketing subroutine \cite{Cho13} to find the interval $[L,R]$ containing the maximum.

\vskip 0.2cm
\State Golden ratio bisection subroutine \cite{Cho13} in the interval $[L,R]$ for finding the maximum $\delta^*$. \Comment{Golden ratio bisection}

\end{algorithmic}
\end{algorithm*}

\subsection{Offline and Online Optimizations}
The joint optimization shown in \textbf{Algorithm \ref{outeroptalg}} can be implemented in two different approaches. In the first, one can generate a set of $\textbf{F}$, user device distances $\textbf{d}=\{d_1,d_2,\dots,d_K\}$ and optimize for an average sum-rate and to find an optimal value for antenna spacing given by $\delta^*$. Then we can fix the antenna spacing at $\delta^*$ (design and implement the antenna system to have $\delta^*$ element spacing) and carryout only the optimization in \textbf{Algorithm \ref{inneroptalg}} periodically while in operation. We refer to this approach as {\em offline optimization}. This approach is attractive in terms of low complexity and is more viable in practical operation. However, another more efficient, but high complexity approach is to have a densely populated antenna element array at the base station with a maximum array dimension $D$ and carryout full joint optimization periodically. Then the elements in the dense antenna array are activated (non-activated elements are to be electronically terminated to avoid passive radiation effects) to maintain the determined optimum $\delta^*$, while power allocation, precoder optimization and the subcarrier selection are also carried-out at the same time. We refer to this approach as {\em online optimization}. Due to the fact that this approach considers the instantaneously available Rayleigh fading coefficients and device locations with respect to the base station, it is expected to provide a superior average optimum sum-rate. However,  due to the large number of function evaluations (hence, massive run of \textbf{Algorithm \ref{inneroptalg}}) this is relatively complex, thus can be considered to determine an upper bound for the sum-rate performance. When low-complexity is desired in lieu of superior sum-rate performance, one can opt for the offline optimization.

\section{Complexity Analysis} \label{Complexity}
Recall that the proposed optimization algorithm performs function evaluations during its outer optimization while each of these function evaluations involve a single run of the inner optimization. Inner optimization itself is a alternating iterative process, thus, it is worth to investigate the complexity involved in the offline and online optimization processes.
During each function evaluation we require to calculate $\hat{\textbf{H}}_{eq}$ once, followed by an iterative run of a number of steps listed in Table \ref{comp1}. Table \ref{comp1} also lists the number of mathematical operations required for each of the different steps. Let us denote the average number of iterations required for the inner optimization algorithm to converge by $\mathcal{I}_{IO}$ and the average number of iterations required for the  iterative power allocation step by $\mathcal{I}_{PA}$. Then the overall average number of operations required for the inner optimization is $N_1+\left\{N_2+N_3+N_4+N_5+(N_6+N_7)\mathcal{I}_{PA}\right\}\mathcal{I}_{IO}$.

The outer optimization involves function evaluations, each of which is a single run of the inner optimization algorithm. In global search algorithms it is customary to investigate the complexity in terms of function evaluations, as each of the function evaluations requires a very large number of mathematical operations compared to the non-function evaluations based mathematical operations. Let the pre-fixed number of iterations in the particle swam optimization and the average number of iterations for Swan bracketing and Golden Ratio bisection be $\mathcal{I}_{PS},\mathcal{I}_{SB}$, and $\mathcal{I}_{BS}$, respectively. With the number of particles $\zeta$ in the particle swarm optimization, we require an average of $\zeta(1+\mathcal{I}_{PS})$ function evaluations for the particle swarm optimization. We also require $2+\mathcal{I}_{SB}$ and $\mathcal{I}_{BS}$ function evaluations for bracketing and bisection algorithms, respectively. Thus, we require an average total of $\left[2+\zeta(1+\mathcal{I}_{PS})+\mathcal{I}_{SB}+\mathcal{I}_{BS}\right]\times \left[N_1+\left\{N_2+N_3+N_4+N_5+(N_6+N_7)\mathcal{I}_{PA}\right\}\mathcal{I}_{IO}\right]$ mathematical calculations for the overall online optimization.\par
However, recall that in the offline optimization, we carryout the outer optimization only once and only the inner optimization is run during the operation. Thus, we perform only a single run of the inner optimization which requires only $N_1+\left\{N_2+N_3+N_4+N_5+(N_6+N_7)\mathcal{I}_{PA}\right\}+\mathcal{I}_{IO}$ mathematical operations resulting in a reduced computational complexity by a factor of $\frac{1}{2+\zeta(1+\mathcal{I}_{PS})+\mathcal{I}_{SB}+\mathcal{I}_{BS}}$.

\begin{table*}[ht]
\caption{Number of mathematical operations required for different steps in inner optimization}
\begin{center}
\begin{tabular}{c c c}
\hline 

Step & Number of mathematical operations&\\[0.5ex]
\hline
$\hat{\textbf{H}}_{eq}$ calculation & $N+\frac{2}{3}N^3+(2N-1)NK$ & $\}N_1$ \\
\hline
$\textbf{Z}_{i,k}$ calculation & $\left[\{(2N-1)+(2N^2+1)\}(K-1)+N^2(K-1)\right]K(M_L+M_H)$ & $\}N_2$\\
\hline
$\textbf{Q}_{i,k}$ and $\mathbf{\Delta}_{i,k}$ calculation & $\mathcal{O}(N^4)\times K(M_L+M_H)$ & $\}N_3$\\
\hline
$\dot{\textbf{h}}_{i,k}$ calculation & $\left[2N+(2N-1)N^2+(2N-1)N\right]K(M_L+M_H)$ & $\}N_4$\\
\hline
Eigenvalues of $\dot{\textbf{h}}_{i,k}\dot{\textbf{h}}_{i,k}^H$& $\left[2N-1\right]K(M_L+M_H)$ & $\}N_5$\\
\hline
Power allocation & $2K(M_L+M_H)+\left[K(M_L+M_H)-1\right]$ & $\}N_6$\\
\hline
$\textbf{w}_{i,k}$ update & $2K(M_L+M_H)$ & $\}N_7$\\
\hline
\end{tabular}
\label{comp1}
\end{center}
\end{table*}

\section{Simulation Results and Discussion} \label{Results}
Let us consider two carriers centered at $3.5$GHz and $17.5$GHz for the low-band and high-band operation, respectively. 
Moreover, let the bandwidth of the low-band sub-carriers be $B_L=120$kHz each and the bandwidth of the high-band sub-carriers be $B_H=480$kHz each. The transmit power at the base-station, $P_T$ is equal to $2$ Watts and the antenna element size is $a=0.0025$m.
Since the noise power is proportional to the channel bandwidth, the high-band channel noise power to low-band channel noise power ratio takes the same value as the bandwidth ratio. 
All sub-carriers undergo quasi-static Rayleigh block fading where each element of the Rayleigh fading channel matrix ${\textbf{F}}_i$ is drawn from a complex Gaussian distribution having zero mean $\frac{1}{2}$ variance per dimension.
We also assume that full knowledge of $\{\textbf{F}_i\}^{M_L+M_H}$ is available at both the transmitter and the receivers. 
Without loss of generality we select the window size vector $\mathbf{\mathcal{N}}$ to be having integer entries drawn from a uniform distribution in the range $[1,M_L+M_H]$. 

\begin{figure}[!t]
\begin{center}
\includegraphics[width= 0.43\textwidth]{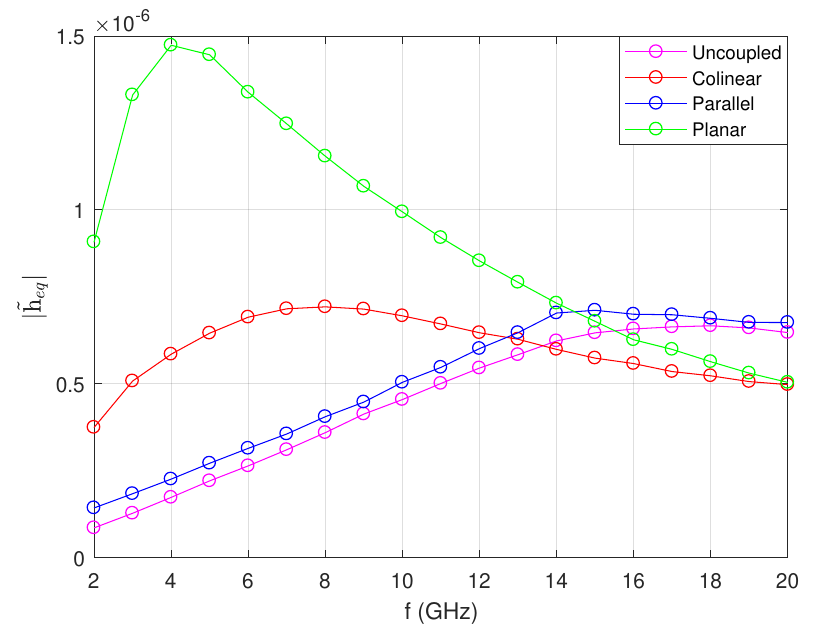}
\end{center}
\caption{Average equivalent channel gain vs. frequency.}
\label{plot1_1}
\end{figure}

We first consider the average equivalent channel gain variation over different frequencies for the optimized antenna spacing and Fig. \ref{plot1_1} shows the magnitude Bode plots for the three antenna configurations considered together with that for a mutually uncoupled transmit antenna array. It is clear that the colinear configuration has a superior wideband response compared to the parallel arrangement. Meanwhile, the rectangular planar array produces the highest response of the three configurations. Parallel configuration produces only a slightly better response than the uncoupled array. 

\begin{figure}[!t]
\begin{center}
\includegraphics[width= 0.43\textwidth]{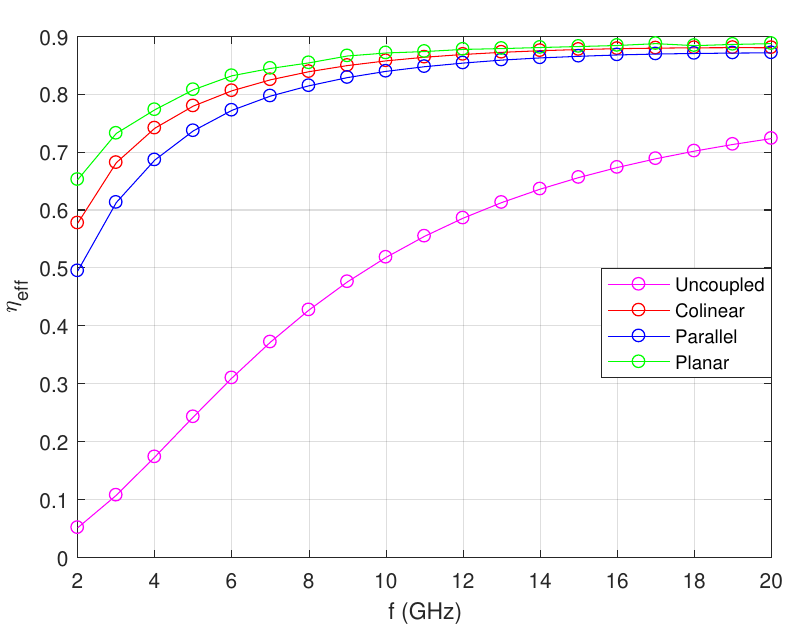}
\end{center}
\caption{Radiation efficiency vs. frequency.}
\label{plot1_2}
\end{figure}

We also show the radiation efficiency against frequency, for the considered antenna configurations in Fig. \ref{plot1_2}. Radiation efficiency variation remains almost the same for the mutual coupled configurations while it is relatively inferior at low frequencies when uncoupled antennas are used.\par
Then we investigate the sum-rate performance of the proposed optimization framework where only the inner optimization is carried out for different BS antenna element spacing values and for different overall array dimensions. Here we perform the inner optimization where power is allocated among all carriers from a common pool. 

\begin{figure}[!t]
\begin{center}
\includegraphics[width= 0.45\textwidth]{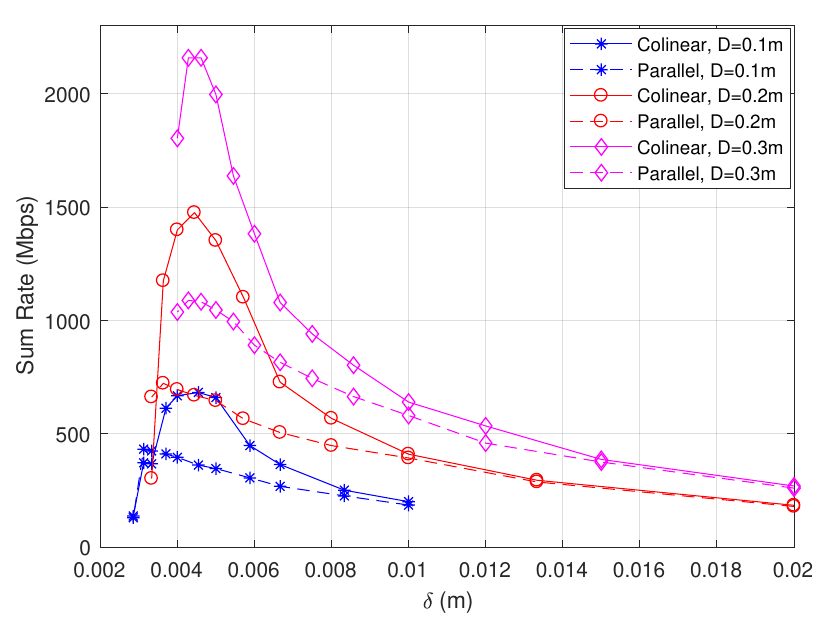}
\end{center}
\caption{Sum-rate vs. BS antenna element spacing for linear BS antenna array.}
\label{plot1}
\end{figure}

\begin{figure}[htbp]
    \centering
    \begin{subfigure}{0.23\textwidth}
        \centering
        \includegraphics[width=1\textwidth]{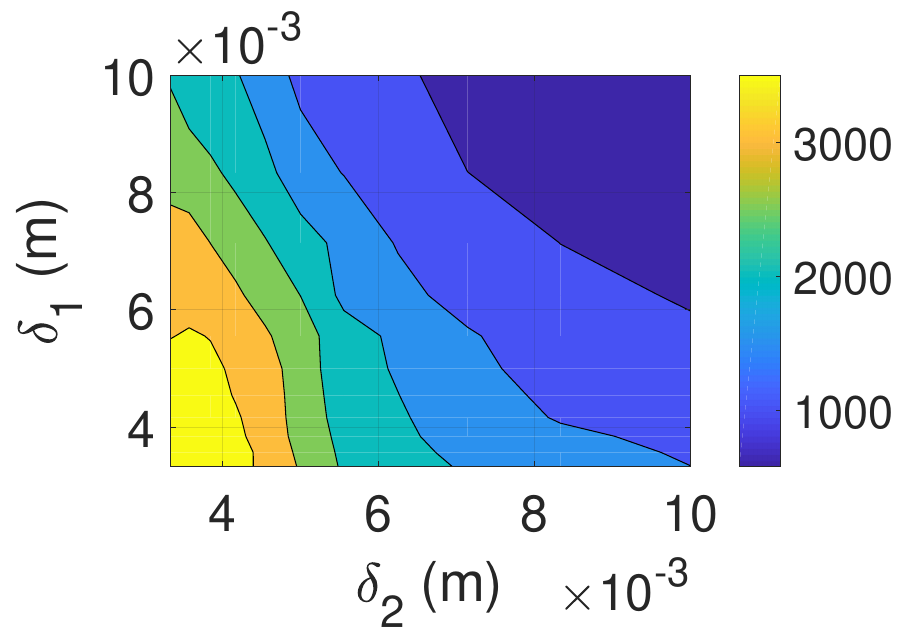}
        (a)
        \label{fig:sub1}
    \end{subfigure}
    \hfill
    \begin{subfigure}{0.21\textwidth}
        \centering
        \includegraphics[width=1\textwidth]{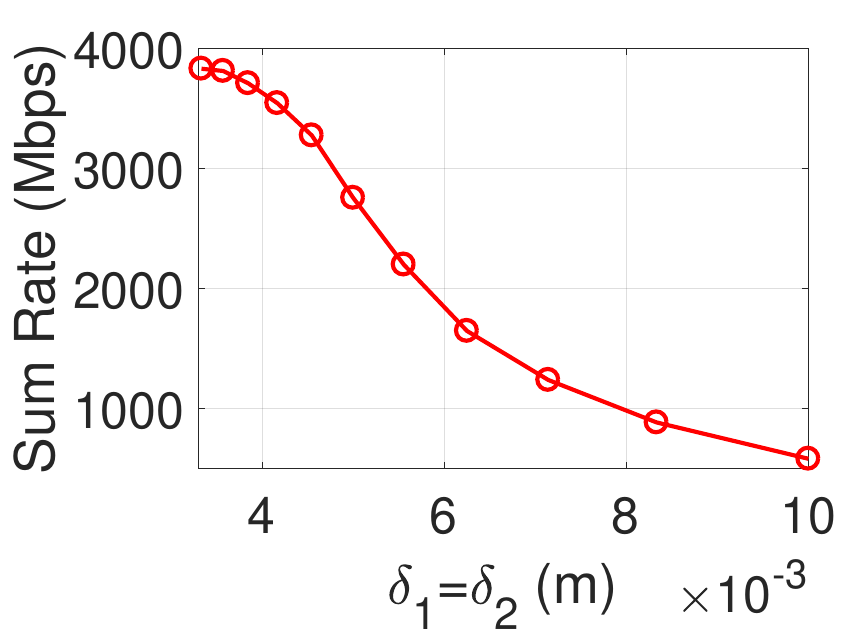}
        (b)
        \label{fig:sub2}
    \end{subfigure}
    \caption{Sum-rate (Mbps) vs. BS antenna element spacing for a 0.05m$\times$0.05m rectangular planar BS antenna array (a) 2D antenna spacing grid (b) Equal vertical and horizontal antenna spacing (i.e., $\delta_1=\delta_2$).}
    \label{PlotPlnar}
\end{figure}

Fig. \ref{plot1} depicts the sum-rate variation with different antenna spacing where the sum-rate improves when the antenna elements are brought closer which is a combined result of the increase of the number of elements resulting in a diversity gain and also the bandwidth gain due to the mutual coupling between antenna elements. However, the sum-rate starts to reduce after a certain optimum spacing value. The reduction is a result of the antenna impedance mismatches and correlations starting to dominate over the aforementioned gains, when the antenna elements are brought closer than the optimum spacing. Furthermore, in Figs. \ref{plot1} and \ref{PlotPlnar} we observe that the optimum sum-rate is higher for colinear array configuration than that for the parallel array configuration. Due to the poor mutual coupling effects existing in the parallel configuration compared to the colinear configuration, the former shows lesser additional bandwidth gain. Thus the resultant optimum sum-rate value becomes  comparatively smaller.

Fig. \ref{PlotPlnar} shows the colour map for the sum-rate of a planar BS antenna array of size $0.05\text{m}\times0.05\text{m}$ vs different antenna element spacing values. It is apparent that the sum-rate performance improves with the reduced antenna spacing and reaches a saturated value, in contrary to the linear arrays where peak sum-rate behaviour is observed. Unlike in the linear arrays, in the planar array antennas in a 2D plane (from different angles) creates mutual coupling effects such that the mutual coupling effects always increases. However, after a certain minimum antenna spacing the spacial correlation takes over the effects of mutual coupling resulting in a saturation behaviour.

Next, we investigate the optimum sum-rate performance for different number of sub-carriers and for different average signal to noise ratio (SNR) values at the receiver antennas. Note that this investigation is an extension of our sum-rate performance comparison in \cite{Bal24} for uncoupled antenna arrays.
We compare the total sum-rate for the
\begin{enumerate}
\item Proposed joint resource optimization under carrier-wise power constraints (i.e., each sub-carrier gets $\frac{P_T}{M_L+M_H}$), denoted as CWPA.
\item Proposed joint resource optimization under band-wise power constraints (i.e., high-band gets $\frac{\beta}{1+\beta} P_T$ and low-band gets $\frac{1}{1+\beta} P_T$, $0\leq \beta < \infty$), denoted as BWPA.
\item Proposed joint resource optimization under full range power constraints (i.e. all carriers are allocated power from a common pool having $P_T$ power), denoted as JPA. We consider $\eta=2$ for all users (Case 1), $\eta=1$ for all users with 75\% users in high-band and 25\% users in low-band  (Case 2), $\eta=1$ for all users with 25\% users in high-band and 75\% users in low-band  (Case 3), and $\eta=2$ for 75\% of the users and $\eta=1$ for 25\% of the users (Case 4).

\end{enumerate}

\begin{figure}[!t]
\begin{center}
\includegraphics[width= 0.48\textwidth]{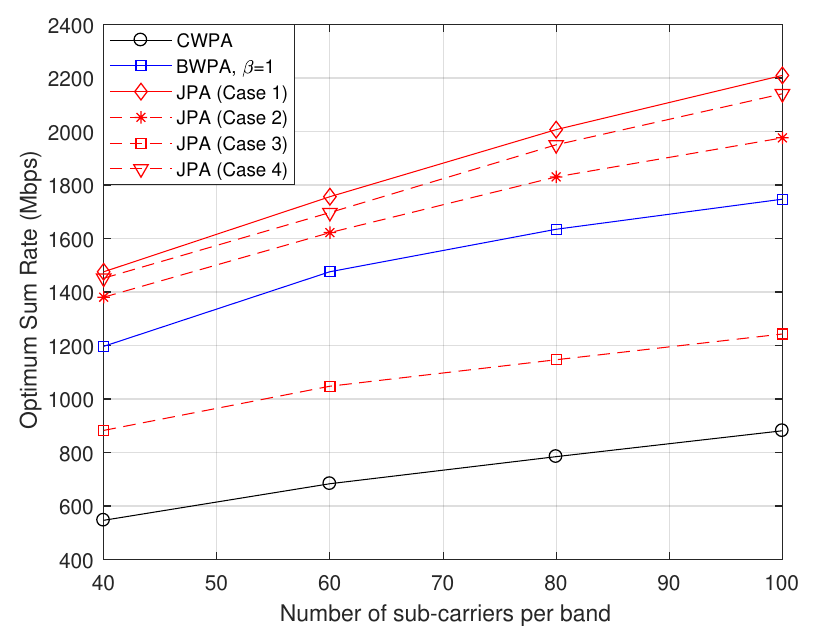}
\end{center}
\caption{Optimum sum-rate vs. number of sub-carriers per band ($D=0.2\text{m}$, colinear array).}
\label{plot3_1}
\end{figure}

\begin{figure}[!t]
\begin{center}
\includegraphics[width= 0.48\textwidth]{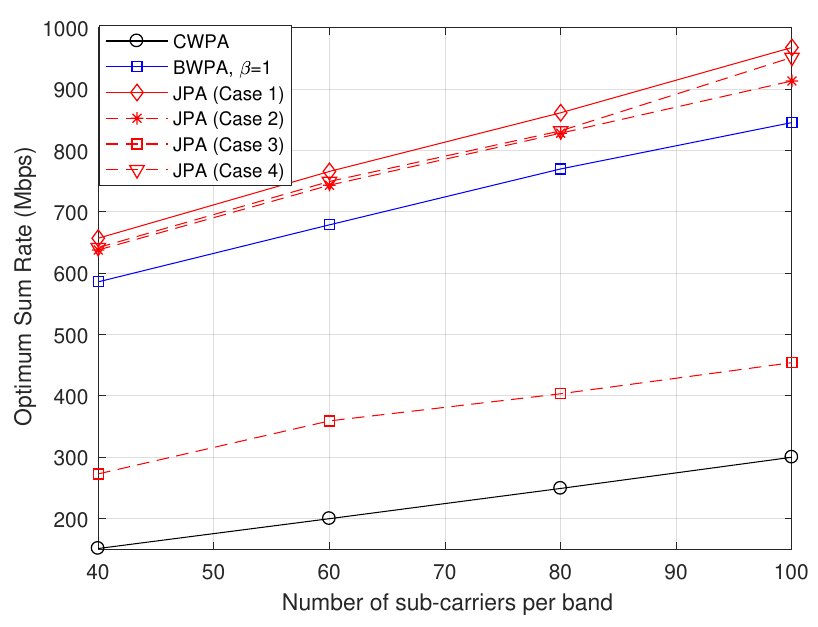}
\end{center}
\caption{Optimum sum-rate vs. number of sub-carriers per band ($D=0.2\text{m}$, parallel array).}
\label{plot3_2}
\end{figure}

\begin{figure}[!t]
\begin{center}
\includegraphics[width= 0.48\textwidth]{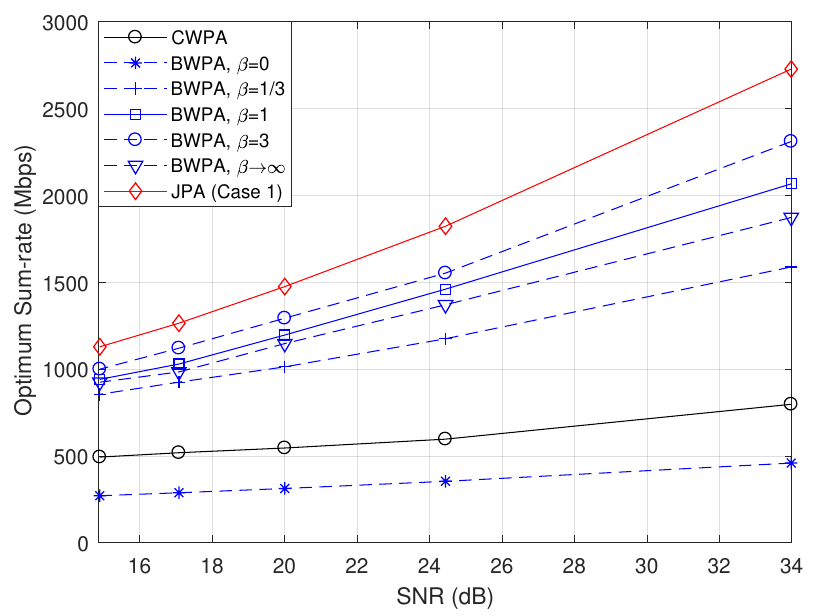}
\end{center}
\caption{Optimum sum-rate vs. SNR ($40$ sub-carriers per band,  $D=0.2\text{m}$, colinear array).}
\label{plot4_1}
\end{figure}

\begin{figure}[!t]
\begin{center}
\includegraphics[width= 0.48\textwidth]{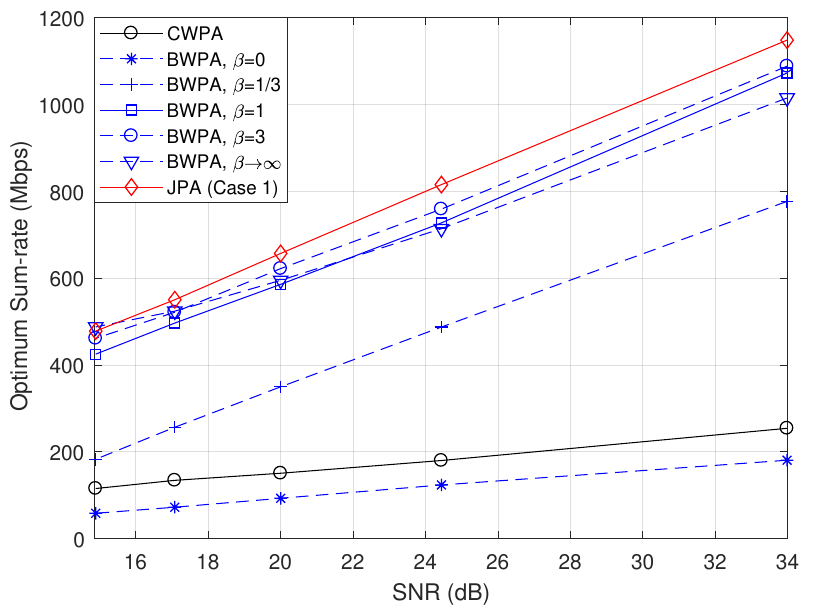}
\end{center}
\caption{Optimum sum-rate vs. SNR ($40$ sub-carriers per band,  $D=0.2\text{m}$, parallel array).}
\label{plot4_2}
\end{figure}

\begin{figure}[!t]
\begin{center}
\includegraphics[width= 0.48\textwidth]{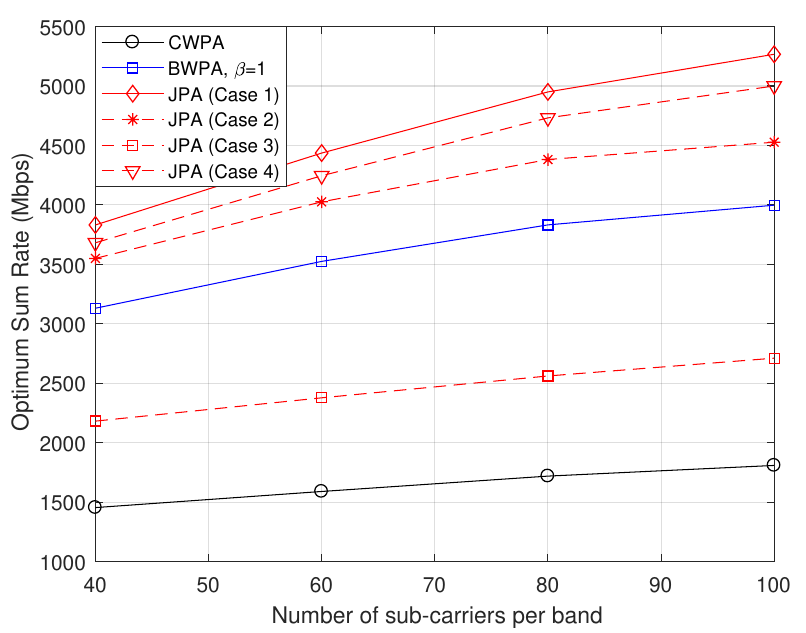}
\end{center}
\caption{Optimum sum-rate vs. number of sub-carriers per band ($0.05\text{m}\times0.05\text{m}$ rectangular planar BS antenna array)}
\label{plot7}
\end{figure}

\begin{figure}[!t]
\begin{center}
\includegraphics[width= 0.48\textwidth]{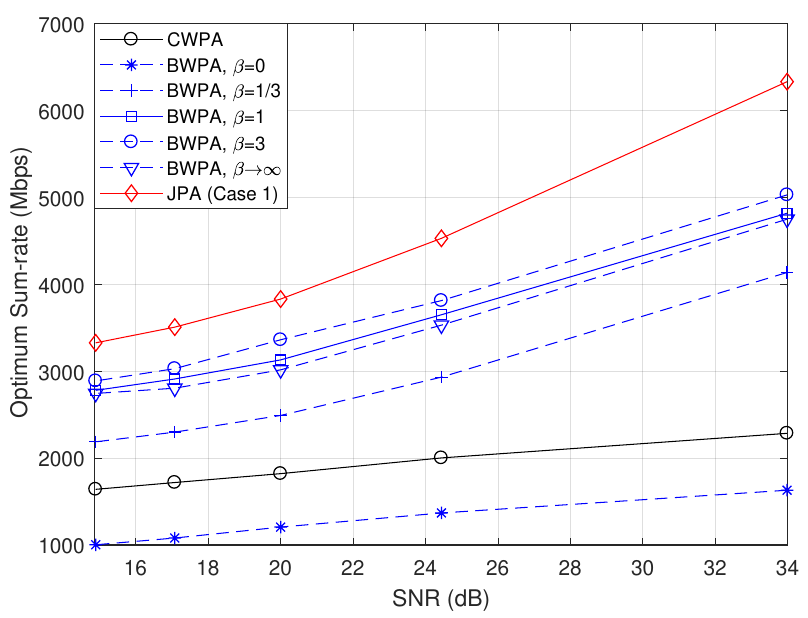}
\end{center}
\caption{Optimum sum-rate vs. SNR ($40$ sub-carriers per band,  $0.05\text{m}\times0.05\text{m}$ rectangular planar array).}
\label{plot8}
\end{figure}

From Figs. \ref{plot3_1}, \ref{plot3_2}, \ref{plot4_1}, and \ref{plot4_2} for linear antenna arrays, it is apparent that the JPA scheme (Case 1) is superior to the CWPA and BWPA ($\beta=1$) schemes resembling the same comparison as in the case of uncoupled arrays. This is due to the fact that, in JPA, carrier-user blocks for allocating power are selected from an overall pool which enables to select the best blocks for higher power allocation. On the contrary, when the power allocation is constrained band wise or sub-carrier wise, it hinders the efficient selection of carrier-user blocks for power allocation. It is also apparent from Figs. \ref{plot3_1} and \ref{plot3_2} that, under all schemes the sum-rate improves with the increased number of sub-carriers. 
Figs. \ref{plot3_1} and \ref{plot3_2} further show that JPA (Case 4) is producing only a marginally inferior sum-rate compared to JPA (Case 1) and JPA (Case 2), whereas, JPA (Case 3) produces a significantly less sum-rate. The higher sum-rate in Cases 1, 2 and 4 are due to the larger unrestricted carrier space to select the sub-carriers than those in Case 3. 

Moreover, from Figs. \ref{plot4_1} and \ref{plot4_2} it is apparent that when the the power allocation ratio, $\beta$ is increased, the sum-rate performance also increases at all the SNR values and reaches a maximum. However, the maximum sum-rate achievable via band-wise power allocation is significantly less than the sum-rate which can be achieved via the JPA scheme. This also implies that the proposed joint power allocation and the associated optimization framework over multiple bands can harness a considerable additional sum-rate than which can be achieved using only the high-band ($\beta \rightarrow \infty$) or the low-band ($\beta=0$).\par
The sum-rate performances at all instances when deploying colinear arrays are superior over the cases using parallel arrays. This is a direct result of the bandwidth gain existing in colinear tightly coupled BS array configurations. 
In Figs. \ref{plot7} and \ref{plot8} we also show the optimum sum-rate performance for different number of carriers per frequency band and also for different SNR values. Although the sum-rate variation is different to that of a linear array, planar array shows a similar behaviour in sum-rate performance over different number of sub-carriers per band and for different SNRs. Joint power allocation where all the users are allowed to allocate power from a common pool, without any restriction outperform all the other power/user allocation schemes.\par
\begin{figure}[!t]
\begin{center}
\includegraphics[width= 0.48\textwidth]{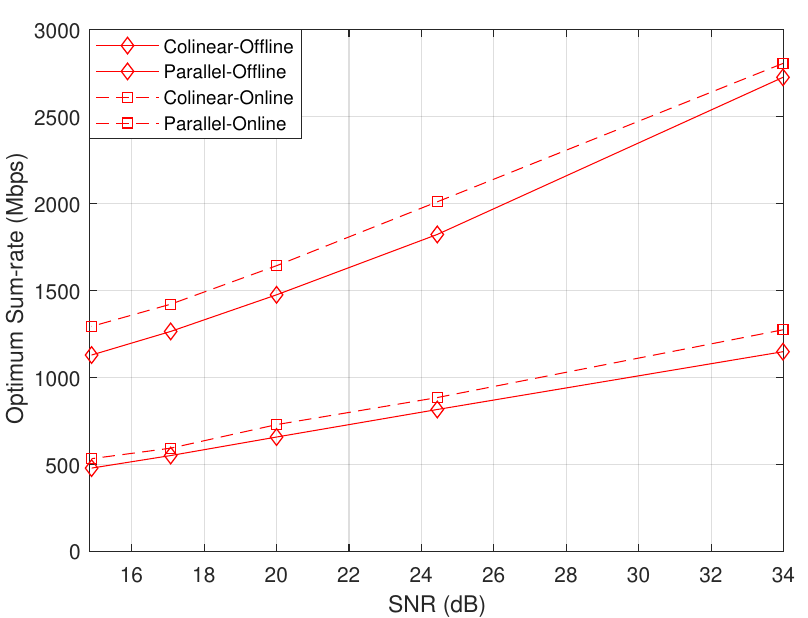}
\end{center}
\caption{Comparison of offline and online optimizations - Optimum sum-rate vs. SNR ($40$ sub-carriers per band).}
\label{Plot9}
\end{figure}

While generating the results in Figs. \ref{plot3_1}-\ref{plot8}, we perform inner optimization a number of times to find an average $g(\delta)$ during each of the function evaluations. It is obvious that we have optimized for an average sum-rate (offline optimization). However, when the instantaneous Raleigh fading realizations and user position sets are considered each time, followed by an online optimization, we can have a superior average optimum sum-rate in linear arrays. Due to the saturating maximum sum-rate behaviour in planar array, which approximates a continuous aperture at small inter-element spacing, the aforementioned additional sum-rate improvements is absent in planar arrays even when the online optimization is performed. In Fig. \ref{Plot9}, we show this optimal sum-rate performance of the online optimization, obtained at the cost of added complexity as discussed in Section \ref{Complexity}.

\section{Conclusion}\label{Conc}
We have proposed a  novel joint resource optimization framework for a physically-consistent multi-band mMIMO system model where BS antenna spacing selection, precoder optimization, sub-carrier selection and power allocation is carried out simultaneously. Antenna spacing optimization allows to harness the favourable effects of mutual coupling to produce a wideband response in the mMIMO antenna array (even with simple individual antenna elements) which in turn allows a more efficient communication in the multi-band scenario. The proposed antenna spacing selection can be carried out either only once at the design stage or dynamically during the operation. We also have investigated the optimum sum-rate performance (obtained via varying the antenna element spacing) variation for different  power allocation and user selection schemes. Through extensive simulation results we have shown that the proposed framework performs the best with power allocation considering all users in a common pool over all frequency bands and sub-carriers.

\end{document}